\documentclass[preprint,showpacs,preprintnumbers,amsmath,amssymb,floatfix]{revtex4}

\usepackage{graphicx}
\usepackage{dcolumn}
\usepackage{bm}
\usepackage[dvipdfm]{hyperref}

\begin{document}

\preprint{}

\title{Induction of nuclear fission by high-voltage application}

\author{Hirokazu Maruyama}
\email{maruyama@comomo.net}
\affiliation{Maruyama Research Institute, 2024 Shimohatacho, Tarumi, Kobe 655-0861, Japan}


\date{\today}

\begin{abstract}
In nuclear power generation, fissile materials are mainly used. 
For example, $U^{235}$ is fissile and therefore quite essential for use of nuclear energy. 
However, the material $U^{235}$ has very small natural abundance less than 1 \%. 
We should seek possibility of utilizing fissionable materials such as $U^{238}$
because natural abundance of such fissionable materials is generally 
much larger than fissile ones. 

In this paper, we show that thermal neutrons with vanishing kinetic energy can 
induce nuclear fission when high voltage is applied to fissionable materials. 
To obtain this result, we use the liquid-drop model for nuclei. 
Finally, we propose how fissionable materials can be utilized. 

\end{abstract}

\maketitle

\section{Introduction}
Due to rapid economic growth in developing countries, there has been 
anxiety that worldwide energy scarcity will occur in the near future. 
Since crude-oil price has started to rise suddenly, 
more effective utilization of nuclear energy should be considered. 
However, the supplied amount of nuclear fuel materials will be 
inevitably insufficient to meet the rapid increase of the energy demand 
because the estimated amount of deposits is limited. 
We should seek possibility of using other nuclear materials for nuclear 
power generation than fissile ones. 

Fissile materials such as U$^{233}$, U$^{235}$, Pu$^{239}$, and Pu$^{241}$ 
absorb neutrons to form compound nuclei. Excitation energy is the sum of 
kinetic and biding energy of an incoming neutron 
in the considered compound nucleus. 
A neutron with vanishing kinetic energy may induce nuclear fission 
if binding energy of the considered compound nucleus is larger than 
the critical energy. 

In contrast, most of heavy nuclei other than U$^{233}$, U$^{235}$, 
Pu$^{239}$, and Pu$^{241}$ cannot let resulting compound nuclei 
reach critical energy when binding energy of incoming neutron is just supplied. 
That is, incoming neutrons need to have sufficient kinetic energy 
to induce nuclear fission. 
This statement always holds when the target nucleus is composed of 
an even number of nucleons because a nucleus with an even mass number gives 
smaller biding energy for incoming neutrons than a nucleus with an odd mass number. 
Those materials are called ``fissionable" 
that have not so high critical energy and can cause nuclear fission 
even with an incoming neutron having low kinetic energy like $0-10$ MeV. 
U$^{238}$ is an example of such fissionable materials. 

Is there a way to utilize fissionable materials in nuclear power generation? 
Certainly, absorption of neutrons with vanishing kinetic energy to 
fissionable materials such as U$^{238}$ is not sufficient to induce nuclear fission. 
However, there is a possibility that incoming neutrons with vanishing 
kinetic energy can induce nuclear fission if the critical energy is decreased 
by applying other external forces. 

In this paper, we show that fissionable materials can produce nuclear fission 
as a result of absorption of neutrons with vanishing kinetic energy 
if high voltage is applied to materials. 
If voltage is applied to metal, nuclei (positively charged) should 
acquire energy proportional to the voltage. 
The whole metal does not obtain energy because the metal is 
neutral concerning to electric charge. 
An electric current proportional to the applied voltage is induced 
and then the electrons (negatively charged) acquire energy. 
There is no increase of energy because nuclei with the same positive charges 
as the total number of the electrons 
lose the same amount of energy as the electrons acquire. 

According to the above consideration, the nuclei absorb energy 
proportional to applied voltage. As a result, 
the Coulomb energy of the nuclei increases, the surface and Coulomb energy 
tend to be off balance, and then the system becomes unstable energetically. 
The instability of energy means that the nuclei can deform and split easily. 
In this paper, we explain this mechanism based on the liquid-drop model 
of Bohr and Wheeler~\cite{rf:0}. 
Also, we show that the critical energy of energetically unstable nuclei 
decreases and finally takes the same value as the biding energy of neutrons. 

The technique of energy loss and gain has laid the foundation of this subject. 
For example, in Ref.~\cite{rf:1},  Reed has used the technique to obtain 
the spontaneous fission limit on $Z^2/A$ and  succeeded nicely in obtaining 
the same result as Bohr and Wheeler in a simple way acceptable to sophomore students. 

In Sec. \ref{mechanism}, we review the mechanism of nuclear fission. 
In Sec. \ref{highvoltage}, we describe the state of nuclei 
when voltage is applied to material and show results of 
critical voltage and nuclear fission voltage. 
Sec. \ref{summary} is devoted to summary and conclusion.

\section{Mechanism of nuclear fission}
\label{mechanism}
In 1939, Bohr and Wheeler gave the first theoretical treatment of nuclear fission 
based on the liquid-drop model, where they considered that nuclei resembled 
charged liquid drops. 
Since nuclear fission is a complicated phenomenon, perfect theory for this phenomenon 
has not been found. For this reason, the old theory given by Bohr and Wheeler is 
still used in the context of nuclear fission. 

Their theory is as follows: 
If a heavy nucleus like U$^{235}$ is excited as a result of absorption 
of thermal neutrons, surface of the liquid drop starts to vibrate, 
which results in deformation of the nucleus from the original shape. 
Besides the Coulomb energy makes the deformation larger. 
If excitation energy is large enough, the Coulomb energy overcomes 
the surface energy and then the nucleus divides into two or more pieces 
of nuclei, each of which has middle amount of mass. 

We are going to explain the details of their theory below. 
To simplify the problem, we assume that a nucleus is a liquid drop with 
constant density and has a definite surface. 
The total volume of a liquid drop is conserved if we assume that 
nuclear matter has constant density, i.e. incompressible. 
That is, vibration of an excited nucleus deforms only the surface. 
The liquid drop deforms from a sphere keeping an axis of symmetry fixed. 
If we consider the axis of symmetry as the polar axis in the polar coordinate 
system, the radial coordinate after deformation of the surface can be 
represented using the Legendre polynomials as follows: 
\begin{equation}
R(\theta)=R_0 \left[1+\sum_{l=0}^{\infty} \alpha_l P_l (cos \theta)\right], 
\label{eq:eq1}
\end{equation}
where $R_0$ is a radius of the spherical liquid drop before deformation, 
$\alpha_l$ are deformation parameters, and $P_l (cos \theta)$ are 
the Legendre polynomials. 
The condition of volume conservation gives $\alpha_0=0$ and $\alpha_1=0$ 
because the center of mass of a liquid drop does not change. 
Therefore, $R(\theta)$ is expanded as 
\begin{equation}
R(\theta)=R_0(1+\alpha_2 P_2+\alpha_3 P_3+...).
\label{eq:eq2}
\end{equation}
If we assume that the nucleus is incompressible (i.e. volume is conserved), 
the surface energy $E_{\rm S}$ is given by ~\cite{rf:2}
\begin{equation}
E_{\rm S}=E_{\rm S0}\left[1+\frac{2}{5} \alpha_2^2-
\frac{4}{105} \alpha_2^3+...\right]. 
\label{eq:eq3}
\end{equation}
We are going to find a condition that small deformation should 
keep a nucleus energetically stable. 
$E_{\rm S0}$ is the surface energy when there is no deformation. 
\begin{equation}
E_{\rm S0}=a_s(1-\kappa_s I^2)A^{2/3}, 
\label{eq:eq4}
\end{equation}
where $a_s$ and $\kappa_s$ are constant numbers related to surface energy 
in the semiempirical formula given by Weizs\"acker and Bethe. 
Coulomb energy $E_{\rm C}$ is 
\begin{equation}
E_{\rm C} =E_{\rm C0} (1-\frac{1}{5} \alpha_2^2-\frac{4}{105} \alpha_2^3+...),
\label{eq:eq5}
\end{equation}
where 
\begin{equation}
E_{\rm C0} =\frac{3}{5}\frac{Z^2e^2}{4\pi\varepsilon_0}\frac{1}{R_{\rm C}}=a_c \frac{Z^2}{A^{1/3}}.
\label{eq:eq6}
\end{equation}
$\varepsilon_0$ and $R_{\rm C}$ are the vacuum electric constant and 
radius of charged sphere of the nucleus, respectively. 
The conventional fissility parameter $x$ is defined as 
\begin{equation}
x =\frac{E_{\rm C0}}{2E_{\rm S0}}
\label{eq:eq7}
\end{equation}
to represent how easily nucleus fission can happen. 
The total deformation energy $E$ of a nucleus is the sum of 
Eqs. (\ref{eq:eq3}) and (\ref{eq:eq5})
\begin{equation}
E \equiv
E_{\rm S}+E_{\rm C}-E_{\rm S0}-E_{\rm C0}=
E_{\rm S0}\left[\frac{2}{5}(1-x)\alpha_2^2-\frac{4}{105}(1+2x)\alpha_2^3+...\right].
\label{eq:eq8}
\end{equation}
The following condition
\begin{equation}
\frac{\partial \Delta E}{\partial \alpha_2}=
E_{\rm S0}\left[ \frac{4}{5}(1-x)\alpha_2-\frac{4}{35}(1+2x)\alpha_2^2\right]=0
\label{eq:eq9}
\end{equation}
gives the extremals of the third-order polynomial of (\ref{eq:eq8}). 
The equation has two roots: 
\begin{equation}
\alpha_2=0, \quad \frac{7(1-x)}{(1+2x)}.
\label{eq:eq10}
\end{equation}
The former and latter correspond to the spherical local minimum and 
barrier maximum, respectively. 
By substituting the latter solution into Eq. (\ref{eq:eq8}), 
we obtain the barrier maximum 
\begin{equation}
E_{\rm barr}=\frac{98}{15}\frac{(1-x)^3}{(1+2x)^2}E_{\rm S0}.
\label{eq:eq11}
\end{equation}

As defined before, the fissility parameter $x$ represents 
how easily nucleus can split. 
Myers and Swiatecki showed that the following equation held 
with $a_c=0.7053$ ($R_c=1.2249 A^{1/3}fm$), $a_s=17.944$, and $\kappa_s=1.7826$ 
~\cite{rf:3}. 
\begin{equation}
x=0.01965\frac{Z^2}{A}\frac{1}{(1-1.7826I^2)}, 
\label{eq:eq12}
\end{equation}
where $I=(N-Z)/(N+Z)$. 

Next, we explain the barrier maximum and biding energy of neutrons, which are both 
necessary for deciding if absorption of a neutron with vanishing kinetic energy 
causes nucleus fission or not. 
When a neutron is absorbed to a nucleus and then a compound nucleus is formed, 
excitation energy $E_{\rm exc}$ is the sum of kinetic $E$ and biding energy 
$E_{\rm b}$ of the incoming neutron contained in the compound system. 
If the biding energy is larger than the barrier maximum $E_{\rm barr}$, 
even neutrons with vanishing kinetic energy can cause nucleus fission. 

Therefore, the condition to cause nucleus fission of the compound 
nucleus is given by 
\begin{eqnarray}
E_{\rm exc}=E_{\rm b}+E \geq E_{\rm barr}.
\label{eq:eq13}
\end{eqnarray}
When an incoming neutron has vanishing kinetic energy, the balance between 
kinetic energy of the neutron and the barrier maximum determines 
if nucleus fission is induced or not. 

\section{Nuclear material in high voltage}
\label{highvoltage}
In the previous section, we have shown the mechanism of nucleus fission 
caused by absorption of neutrons without assuming any other effects. 
In the below, we are going to consider a case when high voltage is applied. 

How are nuclei deformed in metal if high voltage is applied? 
Application of voltage produces an electric current proportional to 
the applied voltage and then electrons (negatively charged) acquire energy. 
Also, nuclei (positively charged) should obtain energy 
proportional to the voltage. Since nuclear material is neutral 
concerning to electric charge, energy of the total system is conserved. 
That is, the nuclei lose the same amount of energy as the electrons 
because the absolute value of the total charge is equal 
between the nuclei and electrons. As a result, the total energy 
of the nuclear material is conserved. 

What happens if nuclei acquire energy as a result of application 
of high voltage on metal? In metal, nuclei are regularly ordered 
and bounded from various directions. The force on nuclei is 
strong and not affected by small voltage like 100 Volts. 
However, there is possibility that nuclei deform if sufficiently 
strong force is generated by applying high voltage in one direction. 
When high voltage is applied to metal, nuclei feel force caused by 
the electric field and try to move in that direction. 
In this case, the nuclei bounded from various directions become 
off balance in the direction parallel to the applied electric field. 
This is similar to the case when rain drops fall freely in an almost 
uniform gravitational field on the earth, where the rain drops deform 
due to air resistance. 

Therefore,  it is expected that nuclei deform if high voltage 
is applied to metal. 
Seeing a thing from a different angle, nuclei are energetically balanced 
between Coulomb force and surface tension if there is no external force. 
However, the Coulomb force becomes dominant over the surface tension 
and nuclei become off  balance if  high voltage is applied. 

Here, we have a question: What is the equation that describes deformation 
of nuclei when high voltage is applied to metal? 
When there is no external force on nuclei, 
the total deformation energy of a nucleus is given by Eq. (\ref{eq:eq8}), 
where the second solution for $\alpha_2$  in Eq. (\ref{eq:eq10}) gives the barrier maximum. 
On the other hand,  force applied to nuclei  in one direction affects on 
the value of $\alpha_2P_2(\theta)$. 
Therefore, the second solution in Eq. (\ref{eq:eq10}) changes 
if high voltage is applied to metal. 
In this case, we should note the following two facts:
\begin{enumerate}
\item Deformation of nuclei is proportional to applied voltage and input electric power 
becomes equal to the barrier maximum of the nuclei in the large-voltage limit. 
\item Nuclei tend to deform and become energetically unstable as $\alpha_2$ decreases. 
Finally, the nuclei split completely at $\alpha_2=0$. 
\end{enumerate}
Based on these facts, it is reasonable to rewrite the second solution of Eq. (\ref{eq:eq10})  to
\begin{equation}
\alpha_2=\frac{7(1-x)}{(1+2x)}-\frac{ZeV}{E_{\rm barr}},
\label{eq:eq16}
\end{equation}
where $Z$ is the number of protons, $e$ is the charge of a proton, 
$V$ is voltage applied, and $E_{\rm barr}$ is the barrier maximum. 

If we substitute this Eq. (\ref{eq:eq16}) into (\ref{eq:eq8}), we have
\begin{equation}
  E_{\rm crit}=\frac{{\left( 686\,E_{S0}\,{\left(x-1 \right) }^4 -
 15\,Z\,e\,V\,{\left( 1 + 2\,x \right) }^3 \right) }^2\,
 \left( 343\,E_{S0}\,{\left(x-1 \right) }^4 + 15\,Z\,e\,V\,{\left( 1 + 2\,x \right) }^3 \right) }
  {24706290\,{\left( 1 - x \right) }^9\,{\left( E_{S0} + 2\,E_{S0}\,x \right) }^2}.
  \label{eq:eq17}
\end{equation}
By specifying the value of $V$ in Eq. (\ref{eq:eq17}), 
we obtain the critical energy of a nucleus when voltage is applied. 

The relation
\begin{eqnarray}
E_{\rm exc}=E_{\rm b}+E \geq E_{\rm crit}
\label{eq:eq13}
\end{eqnarray}
is a condition to cause fission of compound nuclei 
as a result of absorption of neutrons when high voltage is applied.

As shown here, it depends on the balance between the binding energy of 
a neutron and the critical energy of a nucleus with voltage application 
whether nucleus fission is induced or not by neutrons with vanishing kinetic energy.

\subsection{Critical voltage}
If high voltage is applied to fissionable materials like U$^{238}$, 
nuclei becomes unstable and the critical energy decreases 
according to Eq. (\ref{eq:eq17}). 
The same thing applies to compound nuclei that are made as a result of 
absorption of neutrons with vanishing kinetic energy into  fissionable materials. 
If  high voltage with certain strength is applied to compound nuclei, 
the critical energy of a compound nucleus takes the same value as a neutron. 
We have the critical strength of voltage by solving
\begin{equation}
\text{Eq. (\ref{eq:eq17}) = the critical energy of a neutron.} 
\label{eq:eq18}
\end{equation}
We call this the critical voltage. 
The critical energy of a compound nucleus decreases as a result of voltage application 
and then reaches the binding energy of a neutron at the critical voltage. 

When a thermal neutron with almost vanishing kinetic energy is absorbed into U$^{238}$, 
a compound nucleus U$^{239}$ is formed. In this case, U$^{239}$ is in an excited state,  
where the total energy is larger than the ground state energy by the binding energy of a neutron. 
Therefore, if a relation $E_{\rm b} \geq E_{\rm crit}$ holds, absorption of a thermal neutron 
induces nuclear fission. However, if there is no external force, a relation $E_{\rm b} \leq E_{\rm crit}$ 
always holds and nuclear fission is not induced in U$^{239}$. In the above, using the liquid-drop model, 
we have shown that $E_{\rm b} \geq E_{\rm crit}$ holds for U$^{239}$ composed of compound nuclei 
and nuclear fission is induced if the following conditions are met: 
(i) voltage higher than the critical one is applied to the fissionable material U$^{238}$, 
(ii) thermal neutrons are absorbed into the material, and (iii) voltage higher than the critical one  
is continuously applied to the resulting compound nuclei. 

Table \ref{table1} shows the critical voltage for various fissionable materials, 
which has been obtained by substituting  various values  to Eq. (\ref{eq:eq18}). 
For example, Th$^{232}$ gives the values for the compound nucleus Th$^{233}$. 
Since there is no theoretical values for the binding energy of a neutron, 
we have used experimental values alternatively~\cite{rf:4}. 
As shown in the  table, larger difference between the critical and binding energy 
gives  larger critical voltage.

\begin{table}[!h]
\begin{center}
\begin{tabular}{ccccc} \hline\hline
Target nuclei & Compound nuclei & Critical energy &
Neutron's binding energy & Critical voltage \\ \hline
Th$^{232}$ & Th$^{233}$ & 9.7 & 4.8 & $3.8 \times 10^4$ \\
Th$^{233}$ & Th$^{234}$ & 9.9 & 6.2 & $3.1 \times 10^4$ \\
U$^{236}$ & U$^{237}$ & 7.7 & 5.1 & $2.0 \times 10^4$ \\
U$^{238}$ & U$^{239}$ & 7.9 & 4.8 & $2.3 \times 10^4$ \\
U$^{239}$ & U$^{240}$ & 8.0 & 5.9 & $1.9 \times 10^4$ \\
\hline\hline
\end{tabular}
\end{center}
\caption{Estimated values of the critical voltage for fissionable materials. 
Energy and voltage are represented in units of MeV and V, respectively. }
\label{table1}
\end{table}

\noindent
\subsection{Nuclear fission voltage}
As voltage applied to a material increases, nuclei tend to be unstable decreasing the critical energy. 
Finally, the critical energy vanishes. 
The voltage at which the critical energy vanishes is obtained by solving 
\begin{equation}
\text{Eq. (\ref{eq:eq17}) = 0} . 
\label{eq:eq19}
\end{equation}
Hereafter, the voltage is called the nuclear fission voltage. 
Although we need to check some experimental results to identify what physical quantity 
the nuclear fission voltage corresponds to,  we can say that just applying high voltage 
induces  nuclear fission without absorption of neutrons to nuclei. 

Table \ref{table2} shows the nuclear fission voltage, 
which has been obtained  by solving Eq. (\ref{eq:eq19})  for some fissionable materials. 
As shown in the table, the nuclear fission voltage tends to be large as the critical energy increases. 
\begin{table}[!h]
\begin{center}
\begin{tabular}{ccc} \hline\hline
Nuclei  & Critical energy  & Nuclear fission voltage \\ \hline
Th$^{232}$ & 9.6 &  $7.3 \times 10^4$ \\
Th$^{233}$ & 9.7 &  $7.5 \times 10^4$ \\
U$^{236}$ & 7.5 &  $5.2 \times 10^4$ \\
U$^{238}$ & 7.8 &  $5.4 \times 10^4$ \\
U$^{239}$ & 7.9 &  $5.5 \times 10^4$ \\
 \hline\hline
\end{tabular}
\end{center}
\caption{Estimated values of the nuclear fission voltage for fissionable materials. 
Energy and voltage are represented in units of MeV and V, respectively.}
\label{table2}
\end{table}

In Table \ref{table3}, we give the nuclear fission voltage for gold, silver, and copper 
as solutions to Eq. (\ref{eq:eq19}) for reference. 
Clearly, the nuclear fission voltage of these materials is much larger than the fissionable ones. 
\begin{table}[!h]
\begin{center}
\begin{tabular}{ccc} \hline\hline
Nuclei & Critical energy  & Nuclear fission voltage \\ \hline
Cu$^{64}$ & 320 &  $3.7 \times 10^7$ \\
Ag$^{107}$ & 150 &  $7.2 \times 10^6$ \\
Au$^{197}$ & 24 &  $3.1 \times 10^5$ \\
 \hline\hline
\end{tabular}
\end{center}
\caption{Estimated values of the nuclear fission voltage for gold, silver, and copper. 
Energy and voltage are represented in units of MeV and V, respectively.}
\label{table3}
\end{table}

\section{Summary and conclusion}
\label{summary}
We have shown that application of high voltage to fissionable materials such as U$^{238}$ 
induces nuclear fission because the critical energy is decreased to the biding energy of a neutron. 
We have obtained the conditions to induce  nuclear fission based on the liquid-drop model. 

The results obtained in this paper always hold within the given model calculations, 
if all energy injected through voltage application  is used for deformation of nuclei. 
However, in addition to deformation energy of nuclei, 
energy of applied voltage is used for kinetic and vibration energy of nuclei, 
biding energy among nuclei, and excitation energy associated with angular momentum. 
It is difficult to give theoretical estimation for how much amount of the injected energy 
is used for deformation of nuclei. 

For this reason, we need to conduct an experiment to measure if $\gamma$ rays are 
emitted or not after nuclei deformed by voltage application go back to the ground state. 
By conducting such experiments, we can set up a standard concerning to how much 
amount of the injected energy is used for each energy item. 

If emission of $\gamma$ rays is observed, we should conduct an experiment 
to check if voltage application induces nuclear fission or not. 
As shown in our theoretical analysis, nuclear fission may be induced 
if we assume that the whole energy injected by voltage application is used for 
deformation of nuclei through only the critical and biding energy. 
However, it is not clear if we can extract fission energy from nuclei 
because the above condition is for induction of fission of just one nucleus. 
To induce nuclear fission and trigger chain reaction, the average number of neutrons 
produced by nuclear fission must be two or more and the cross section of nuclear fission 
caused by thermal neutrons must be large. 

It is difficult to check theoretically if these conditions are met or not. 
Also, there is possibility that reaction cross section of a thermal neutron changes 
due to decrease of the critical energy in these materials. 
For this reason,  we should confirm how much voltage needs to be added to the obtained 
critical voltage for inducing nuclear fission based on an experiment in each material. 
In each experiment, we have to count the number of secondary neutrons and 
measure cross section of nuclear fission. 

If all of the conditions are met, application of voltage to fissionable materials such as $U^{238}$ 
induces nuclear fission with absorption of thermal neutrons with vanishing kinetic energy. 
That is, fissionable materials, which have much larger natural abundance than fissile ones, 
can be used as nuclear fuel. 
In this way, we can contribute to the energy problems by utilizing nuclear waste.




\end{document}